\documentclass[10pt, conference, letterpaper]{IEEEtran}

\usepackage{cite}
\usepackage{graphicx,color,epsfig,rotating}
\usepackage{amsfonts,amsmath,amssymb,bbm}
\usepackage{algorithm,algorithmic}
\usepackage{subfigure}
\usepackage{amsmath}
\usepackage{cite}
\usepackage{mdwtab}
\usepackage{placeins}
\usepackage{psfrag, graphicx}
\usepackage[latin1]{inputenc}
\usepackage{amssymb}
\usepackage{multirow}
\usepackage{mathrsfs}

\long\def\comment#1{}

\newfont{\bbb}{msbm10 scaled 700}

\newfont{\bb}{msbm10 scaled 1100}

\newcommand{\PP}{\mbox{\bb P}}

\newcommand{\FF}{\mbox{\bb F}}

\newcommand{\EE}{\mbox{\bb E}}

\newcommand{\cv}{{\bf c}}

\newcommand{\fv}{{\bf f}}

\newcommand{\Cm}{{\bf C}}

\newcommand{\Pm}{{\bf P}}
\newcommand{\Qm}{{\bf Q}}

\newcommand{\Wm}{{\bf W}}

\newcommand{\Cc}{{\cal C}}

\newcommand{\Ec}{{\cal E}}
\newcommand{\Fc}{{\cal F}}
\newcommand{\Gc}{{\cal G}}
\newcommand{\Hc}{{\cal H}}
\newcommand{\Ic}{{\cal I}}

\newcommand{\Kc}{{\cal K}}
\newcommand{\Lc}{{\cal L}}

\newcommand{\Qc}{{\cal Q}}

\newcommand{\Uc}{{\cal U}}
\newcommand{\Wc}{{\cal W}}
\newcommand{\Vc}{{\cal V}}

\newcommand{\Csf}{{\sf C}}

\newcommand{\Hsf}{{\sf H}}

\newcommand{\Wsf}{{\sf W}}

\renewcommand{\arg}{{\hbox{arg}}}

\newcommand{\eqdef}{\stackrel{\Delta}{=}}

\newcommand{\be}{\begin{equation}}
\newcommand{\ee}{\end{equation}}
\newcommand{\bea}{\begin{eqnarray}}
\newcommand{\eea}{\end{eqnarray}}

\newtheorem{theorem}{Theorem}%[section]
\newcommand{\ie}{{\it i.e., }}

\begin{document}

\sloppy

%% Paper Title
%% You can use linebreaks \\ within to get better formatting as
%% desired.

%\title{Caching-Aided Coded Multicast for Efficient Content Delivery with Popularity Distribution}
%\title{An Efficient Coloring Algorithm for Caching-Aided Coded Multicasting}
\title{An Efficient Coded Multicasting Scheme Preserving the Multiplicative Caching Gain}

%% Author names and affiliations:
%%
%% Avoiding spaces at the end of the author lines is not a problem with
%% conference papers because we don't use \thanks or \IEEEmembership.
%%
%% For several authors with only one affiliation:
%%
% \author{
%   \IEEEauthorblockN{Hui-Ting Chang and Stefan M.~Moser}
%   \IEEEauthorblockA{Department of Electrical and Computer Engineering\\
%     National Chiao Tung University (NCTU)\\
%     Hsinchu, Taiwan\\
%     Email: \{email-of-hui-ting,email-of-stefan\}@ieee.org}
% }
%%
%% For up to three affiliations:
%%
%%
%% For over three affiliations, or if they all won't fit within the width
%% of the page, use this alternative format:
%%%
% \author{
%   \IEEEauthorblockN{
%     Giuseppe Vettigli \IEEEauthorrefmark{1},
%     Mingyue Ji \IEEEauthorrefmark{2},
%     Antonia Tulino \IEEEauthorrefmark{3},
%     Jaime Llorca \IEEEauthorrefmark{3} and
%     Paola Festa \IEEEauthorrefmark{1}}
%   \IEEEauthorblockA{
%     \IEEEauthorrefmark{1}School of Electrical and Computer Engineering\\
%     University of Napoli Federico II \\
%     Email: }
%   \IEEEauthorblockA{
%     \IEEEauthorrefmark{2}Alcatel-Lucent, Bell Labs, USA}
%   \IEEEauthorblockA{
%     \IEEEauthorrefmark{3} EE department, University of Southern California \\
%     Email: mingyuej@usc.edu}
%   \IEEEauthorblockA{
%     \IEEEauthorrefmark{4}Tyrell Inc., 123 Replicant Street, Los Angeles, California 90210--4321}
% }

 \author{
   \IEEEauthorblockN{
     Giuseppe Vettigli\IEEEauthorrefmark{1},
     Mingyue Ji\IEEEauthorrefmark{2},
     Antonia M. Tulino\IEEEauthorrefmark{1}\IEEEauthorrefmark{3},
     Jaime Llorca\IEEEauthorrefmark{3}, 
     Paola Festa\IEEEauthorrefmark{1} 
   %  Eldon Tyrell\IEEEauthorrefmark{4}
    }
   \IEEEauthorblockA{
     \IEEEauthorrefmark{1}Universit\'a di Napoli Federico II, Napoli, Italy.     Email: \{vettigli, festa\}@unina.it}
 \IEEEauthorblockA{
     \IEEEauthorrefmark{2}EE Department, University of Southern California, CA.
     Email: mingyuej@usc.edu}
   \IEEEauthorblockA{
     \IEEEauthorrefmark{3}Alcatel Lucent, Bell labs, NJ.
     Email: \{a.tulino, jaime.llorca\}@alcatel-lucent.com}
%     Telephone: (800) 555--1212, Fax: (888) 555--1212}
%   \IEEEauthorblockA{
%     \IEEEauthorrefmark{4}Tyrell Inc., 123 Replicant Street, Los Angeles, California 90210--4321}
 }

%% Use for special paper notices

%% To balance the two columns, you should reduce the text-height of
%% the last page using the following command:
%%%%%%%%%%%%%%%%%%%%%%%%%%%%%%%%%%%%%%%%%%%%%%%%%%%%%%%%%%%%%%%%%%%%%
%\addtolength{\textheight}{-9.35cm}
%%%%%%%%%%%%%%%%%%%%%%%%%%%%%%%%%%%%%%%%%%%%%%%%%%%%%%%%%%%%%%%%%%%%%
%% with an appropriate value. This command must be place on the second
%% last page, i.e., for a one-page abstract here, for a two-page
%% abstract right after the \maketitle command.

%% Create the title:
\maketitle

\begin{abstract}

Coded multicasting has been shown to be a promising approach to significantly improve the caching performance of content delivery networks with multiple caches downstream of a common multicast link. 
However, achievable schemes proposed to date have been shown to achieve the proved order-optimal performance only in the asymptotic regime in which the number of packets per requested item goes to infinity.
In this paper, we first extend the asymptotic analysis of the 
achievable scheme in \cite{ji2014average,ji2015orderoptimal} 
to the case of heterogeneous cache sizes and demand distributions, providing the best known upper bound on the fundamental limiting performance when the number of packets goes to infinity. We then show that the scheme achieving this upper bound quickly loses its multiplicative caching gain for finite content packetization. To overcome this limitation, we design a novel polynomial-time algorithm based on random greedy graph-coloring that, while keeping the same finite content packetization, recovers a significant part of the multiplicative caching gain. Our results show that the order-optimal coded multicasting schemes proposed to date, while useful in quantifying the fundamental limiting performance, must be properly designed for practical regimes of finite packetization.

\end{abstract}

%\vspace{-0.2cm}
\section{Introduction}
\label{section: intro}

%In \cite{llorcatulino132, llorcatulino14, ji2013optimalJ, ji2013wireless, ji2013fundamental, ji2014average, ji2014multiple, shanmugam2013femtocaching, maddah2012fundamental, maddah2013decentralized}, caching has been shown to be a potential technique to solve the problem of the exponentially increased data traffic in today's network, which is mainly driven by the service of on-demand video such as Netflix and Amazon Prime \cite{cisco13}.

Recent studies \cite{llorcatulino132, ji2013optimalJ, ji2014average, ji2015orderoptimal, ji2013wireless, maddah2012fundamental, maddah2013decentralized, ji2014fundamental,ji2014multiple,Shanmugam2014Finite} have been able to characterize the information theoretic limiting performance of several caching networks of practical relevance, in which throughput scales linearly with cache size, showing great promise to accommodate the exponential traffic growth experienced in today's communication networks \cite{cisco13}. %{\RED check refs and order them!}

Consider a network with one source (server), having access to $m$ files, and $n$ users (caches), each with a storage capacity of $M$ files.
%Let a network consisting of a source node (base station), $n$ users, $m$ files and each user node has storage capacity of $M$. 
In \cite{ji2013optimalJ}, % ji2013wireless
the authors showed that if the users can communicate between each other via Device-to-Device (D2D) communications, %are allowed and without using the source node, for an arbitrarily small outage probability, \footnote{Outage event is defined as that users are not served by D2D links.} the authors present 
a simple distributed random caching scheme and TDMA-based unicast D2D delivery achieves the order-optimal\footnote{Order-optimal means that the gap between the information theoretic converse and the achievable throughput can be bounded by a constant number when $m,n \rightarrow \infty$.} 
worst-case throughput law $\Theta\left(\max\{\frac{M}{m}, \frac{1}{m}, \frac{1}{n}\}\right)$,%\footnote{Given two functions $f$ and $g$, we say that: 1)  $f(n) = O\left(g(n)\right)$ if there exists a constant $c$ and integer $N$ such that  $f(n)\leq cg(n)$ for $n>N$; 2) $f(n) = \Theta\left(g(n)\right)$ if $f(n) = O\left(g(n)\right)$ and $g(n) = O\left(f(n)\right)$.} %without even using the shared link during delivery. The 
whose linear scaling with $M$ when $Mn \geq m$ exhibits a remarkable multiplicative caching gain. Moreover, in this scheme each user caches entire files without the need of partitioning files into packets, %2) $f(n)=o\left(g(n)\right)$ if $\lim_{n \rightarrow \infty}\frac{f(n)}{g(n)} = 0$. 
%2) $f(n) = \Omega\left(g(n)\right)$ if $g(n) = O\left(f(n)\right)$. %4) 
%$f(n) = \omega\left(g(n)\right)$ if $g(n) = o\left(f(n)\right)$. 
and missed files are delivered via unicast transmissions between neighbor nodes, making it efficiently implementable in practice. 

In the case that users cannot communicate between each other, but share a multicast link from the content source, %(e.g., users cannot serve each other due to power constraints) %(e.g., a wireline network where users have no connection to each other),
%for a network with a shared multicast link, 
the authors in \cite{maddah2012fundamental, maddah2013decentralized} showed that the use of coded multicasting allows achieving  
%the authors in \cite{maddah2012fundamental, maddah2013decentralized} presented a deterministic caching and coded multicasting scheme achieving 
the same order-optimal worst-case throughput as in the D2D caching network, 
proving the remarkable fact that multiplicative caching gains can be preserved even if caches cannot communicate between each other. 
However, the linear coding schemes in \cite{maddah2012fundamental, maddah2013decentralized} involve a number of computations that can grow exponentially with the number of users (caches) and the proved order-optimal performance is only guaranteed when
%requires a centralized caching policy and each file to be partitioned into a number of packets that also grows exponentially with the number of users
the number of packets per requested file either goes to infinity or also grows exponentially with the number of users. %, significantly limiting their practical implementation. 
%For example, in a cellular network such as a university campus, where there is one base station, $10000$ users, $300$ files (e.g., top $300$ popular movies in Netflix) and storage capacity $20$ files each user, then it needs ${10000 \choose 600} > 10^{15}$ packets per file, highly infeasible in practice {\RED [relax this a bit]}. 
%In \cite{maddah2013decentralized}, the authors presented an alternative scheme for the same network that uses a simpler decentralized random caching policy while a more complex coded multicasting scheme requiring a number of computations that grows exponentially with the number of users. Nonetheless, to guarantee the same throughput, the file size (or equivalently the number of packets per file) is required to go to infinity. 

In \cite{ji2014average}, the authors considered the same shared link network under random demands characterized by a demand distribution, and proposed a scheme consisting of a random popularity-based (RAP) caching policy and a chromatic-number index coding (CIC) based multicasting scheme, referred to as RAP-CIC, proved to be order-optimal in terms of average throughput. The authors further provided the optimal scaling laws for all regions of the Zipf \cite{breslau1999web} demand distribution, whose analytical characterization required resorting to a polynomial-time approximation of CIC, referred to as greedy constrained coloring (GCC). While GCC exhibits polynomial complexity in the number of users and packets, the order-optimal performance guarantee %as with CIC and the schemes in \cite{maddah2012fundamental, maddah2013decentralized}, 
 still requires the number of packets per file to go to infinity.

%{\RED (something about multiplicative caching gain for all regions, etc)} In order to analytically quantify the performance of RAP-CIC,  the authors in \cite{ji2014average} resorted to a polynomial-time approximation of CIC, referred as greedy constrained coloring (GCC) that guarantees the order-optimal throughput when the number of packets per file goes to infinity. {\RED RAP-GCC is also shown to achieve the same performance as the algorithm in\cite{maddah2013decentralized} for the worst-case demand setting. }

It is then key to understand if using any of above referenced schemes, the promising linear throughput scaling with cache size (multiplicative caching gain) can be preserved in practical settings of finite file packetization.   
In this paper, we address this important open problem focusing  on a 
non-homogenous caching network with a shared multicast link, where users make requests
according to possibly different demand distributions and have
possibly different cache sizes. %(see Fig.~\ref{fig:
%network_overview} as an example). 
%The considered network is a generalization of the one considered in \cite{ji2014average},  where all the users are assumed to have the same demand distribution and equal storage capacity. 
The contributions of this paper are as follows.
First, we extend RAP-CIC and RAP-GCC to the non-homogenous shared link network and quantify their average performance. 
%Next,  we extend the polynomial-time algorithm proposed in \cite{ji2014average} for designing the coded multicast transmission scheme providing an upper bound of the RAP average achievable rate   in terms of the average number of equivalent file transmissions (inversely proportional to the throughput) and based on such upper-bound we develop  an optimization framework for the caching policy, which is the first time shown in literatures according to our knowledge. 
Next, we focus on finite file packetization regimes and numerically show that neither GCC nor the coded delivery schemes in \cite{maddah2012fundamental, maddah2013decentralized}
can guarantee the promising performance. %for finite file packetization.
Finally, we introduce 
%two polynomial-time algorithms, where the first
%algorithm can achieve the upper bound of the achievable rate given
%by this paper, and the second 
a novel algorithm based on a greedy randomized approach referred to as Greedy Randomized Algorithm Search Procedure (GRASP), which is shown to recover a significant part of the multiplicative caching gain in the same regimes of finite file packetization,
while incurring a complexity at most quadratic in the number of requested packets. 

%The caching caching by using the proposed algorithm under finite file size is verified by extensively practical simulations.

%The paper is organized as follows. Section \ref{section: network model} introduces the network model and problem formulation. The achievable caching and coded delivery scheme, along with the general upper bound of the average achievable rate are presented in Section \ref{sec: Achievable Delivery Scheme}. Section \ref{sec: algorithms} describes
%the proposed polynomial-time coloring algorithm. Finally, Section \ref{sec: Simulations and Discussions} presents the simulation results and related discussions, and we conclude the paper in Section \ref{sec: Conclusion}.

%\begin{figure}[ht]
%\centerline{\includegraphics[width=6cm]{network_overview}}
%\caption{An example of the network model, which consists of a source node (base station in this figure) having access to the content library and connected to the users via a shared bottleneck (multicast) link. Each user may have different cache size and request files according to its own demand distribution.}
%\label{fig: network_overview}
%\end{figure}

%%%%%%%%%%%%%%%%%%%%%%%%%%%%%%%%%%%%%%%%%%%%%%%%%%%%%%
%\vspace{-0.2cm}
\section{Network Model and Problem Formulation}
\label{section: network model}

We consider a network consisting of a source node with access to a content library of $m$ files $\Fc=\{1,\dots, m\}$ each of size $F$ bits, and
 $n$ user nodes $\Uc=\{1,\dots, n\}$. The source node communicates with the user nodes through a shared
multicast link of finite capacity $C$. Without loss of generality, we assume $C = F$ bits/unit time
and measure transmission rate in time units or file-size transmissions necessary to deliver the requested files to the users.
Each user $u \in \Uc$ has a cache of size $M_uF$ bits (\ie $M_u$ files).
The channel between the source and all the users follows a shared error-free deterministic model.
User $u$ requests file $f$ with probability $q_{f,u}$, where $q_{f,u} \in [0,1]$ and $\sum_{f=1}^m q_{f,u} = 1$. We denote by $\Qm = [q_{f,u}], u=1, \dots, n, f=1, \dots, m$, the demand (or content popularity) distribution, and by $\fv = \{f_1, \dots, f_n\}$ the request vector, where $f_u$ is the file requested by user $u$.
%Let the file requested by user $u$ be $f_u$ such that that the requested vector by all the users can be denoted as $\fv = \{f_1, \dots, f_n\}$. 
%An example of the network model is shown in Fig. \ref{fig: network_overview}.
The goal is to design a content distribution scheme 
consisting of a caching placement (configuration of the user caches) and
a delivery scheme (multicast codeword to be sent to all users) such that all demands are satisfied with probability $1$ and the
expected rate $\bar R(\Qm)$ is minimized. %\footnote{The expected rate is defined as the average minimum number of file transmissions, which is inversely proportional to the average throughput. {\RED consistency with rate definition!!}}  
The expectation is over the demand distribution $\Qm $. We denote the minimum achievable expected rate by $\bar R^*(\Qm)$.

%%%%%%%%%%%%%%%%%%%%%%%%%%%%%%%%%%%%%%%%%%%%%%%%%%%%%%%%%%%
%\vspace{-0.1cm}
%\section{Achievable Scheme}
\section{General Achievable Scheme}
\label{sec: Achievable Delivery Scheme}

%In this section, 
In this section, we extend the analysis of RAP-CIC \cite{ji2014average}, an achievable scheme based on random popularity-based (RAP) caching and chromatic index coding (CIC) delivery, to the heterogeneous shared link network introduced in Section \ref{section: network model}.

%%%%%%%%%%%%%%%%%%%%%%%%%%%%%%%%%%%%%%%%%%%%%%%%%%%
%\vspace{-0.2cm}
\subsection{RAndom Popularity-based (RAP) Caching}
\label{sec: Caching Placement Scheme}

As in \cite{ji2014average, ji2015orderoptimal}, let each file be partitioned into $B$ equal-size packets, represented as symbols of $\FF_{2^{F/B}}$ for $F/B \rightarrow$ constant as $F, B \rightarrow \infty$.
We denote by $\Cm$ and $\Wm$ the realizations of the {\em packet level} caching and demand configurations, respectively, where $\Cm_{u,f}$ denotes the packets of file $f$ cached by user $u$, and $\Wm_{u,f}$ the packets of file $f$ requested by user $u$. 
We use the caching algorithm in Fig.~\ref{alg1} to let each user fill its own cache independently %(and therefore in a decentralized way) 
by knowing the caching distribution $\Pm=[p_{f,u}]$ with $0 \leq M_u p_{f,u}\leq1, \forall f$, and $\sum_{f=1}^m p_{f,u}=1, \forall u$. In line with \cite{ji2014average}, we refer to the caching policy in Fig.~\ref{alg1} that uses the caching distribution that minimizes the upper bound of the optimal expected rate given in Section \ref{sec: achievable} as RAP.

\subsection{Chromatic Index Coding (CIC) Delivery}
\label{sec: Coded Transmission}

%Our coded delivery scheme is based on chromatic number index coding \cite{blasiak2010index, ji2014average}.
%The (undirected) conflict graph $\mathcal H_{\Cm, \Wm} = (\Vc, \Ec)$, where $\Vc$ and $\Ec$ denote the set of vertices and edges of $\mathcal H_{\Cm, \Wm}$, respectively, is constructed as follows:
%\begin{itemize}
%\item Consider each packet requested by each user as a distinct vertex, i.e., if the same packet is requested
%by $N > 1$ users, it results in $N$ distinct vertices.
%\item Create an edge between vertices $v_1, v_2 \in \Vc$ if: 1) they do not represent the same packet,
%and 2) $v_1$ is not available in the cache of the user requesting $v_2$,
%or $v_2$ is not available in the cache of the user requesting $v_1$.
%
%\end{itemize}
The CIC delivery scheme is based on a minimum vertex coloring of the corresponding index coding conflict graph \cite{bar2011index}, $\mathcal H_{\Cm, \Wm} = (\Vc, \Ec)$, constructed as follows: %blasiak2010index
%, oded multicast delivery scheme is based on chromatic number index coding \cite{blasiak2010index}. %\cite{blasiak2010index, ji2014average}. The directed %(undirected) 
%conflict graph $\mathcal H_{\Cm, \Wm} = (\Vc, \Ec)$ %, where $\Vc$ and $\Ec$ denote the set of vertices and edges of $\mathcal H_{\Cm, \Wm}$, respectively, is constructed as follows:
\begin{itemize}
\item Each packet requested by each user is represented as a distinct vertex in $\mathcal V$. %, i.e., if the same packet is requested by $N > 1$ users, it results in $N$ distinct vertices. Hence, each vertex of 
%$\mathcal H_{\Cm, \Wm}=(\Vc, \Ec)$. 
Each vertex $v\in\Vc$ is hence uniquely  identified by a pair 
$\{ \rho(v),\mu(v)\}$  where $\rho(v)$ indicates the  \mbox{packet identity} and 
$\mu(v)$ the \mbox{user requesting it}.
\item For any two vertices $v_1$, $v_2$, we say that vertex $v_1$ interferes with vertex $v_2$ 
if the packet associated with $v_1$, $\rho(v_1)$, is not in the cache of the user associated with  $v_2$, $\mu(v_2)$, 
and $\rho(v_1)$ and $\rho(v_2)$ do not represent the same packet. There exists an undirected edge between $v_1$ and $v_2$ if $v_1$ interferes with $v_2$ or $v_2$ interferes with $v_1$.
%Create an edge between vertices $v_1, v_2 \in \Vc$ if: 1) they do not represent the same packet,
%and 2) $v_1$ is not available in the cache of the user requesting $v_2$,
%or $v_2$ is not available in the cache of the user requesting $v_1$.
\end{itemize}

Given a minimum vertex coloring of the conflict graph $\mathcal H_{\Cm, \Wm}$, %(see Fig.~\ref{fig: conflict_ex_coloring}),
the corresponding index coding scheme transmits the modulo sum of the packets (vertices in $\mathcal H_{\Cm, \Wm}$)
with the same color. Therefore, given $\Cm$ and $\Wm$, the total number of packet transmissions given by
the chromatic number of the conflict graph $\chi(\mathcal H_{\Cm, \Wm})$, which yields a transmission rate of 
$\chi(\mathcal H_{\Cm, \Wm})/B$. 
%In the following, we refer to this coding scheme scheme as CIC (chromatic index coding). 

%\vspace{-0.5cm}

\begin{figure}
\centerline{\fbox{\begin{minipage}[h]{7 cm} \normalsize{
{\bf algorithm} {\tt Caching algorithm ($\Pm$)} \\
$1$\hspace*{0.2truecm} {\bf for} $f \in \mathcal{F}$ \\
%$2$\hspace*{0.7truecm} Each user $u$ randomly caches a subset of \\
%\hspace*{0.8truecm} packets with cardinality of $p_{f,u} M B$  \\
%\hspace*{0.8truecm}  from file $f$ according to $[p_{f,u}]_{f=1}^m$; \\
$2$\hspace*{0.7truecm}Each user $u\in\mathcal U$ caches a subset $\Cm_{u,f}$ of \\
\hspace*{0.8truecm} $p_{f,u} M_u B$ distinct packets of file $f$ \\
\hspace*{0.8truecm} uniformly at random; \\
$3$\hspace*{0.1truecm} {\bf endfor} \\
$4$\hspace*{0.5truecm} $\Cm \leftarrow \{\Cm_{u,f}, u = 1, \dots, n, f = 1, \dots, m\};$ \\
%$\hspace*{2truecm} f = 1, \dots, m\}$; \\
$5$\hspace*{0.0truecm} {\bf return}($\Cm$);          \\
{\bf end} {\tt Caching algorithm} }
\end{minipage}}}
\caption{The random caching algorithm.}
\label{alg1}
\vspace{-0.4cm}
\end{figure}

\subsection{Achievable Expected Rate}
\label{sec: achievable}

%{ \RED Mingyue:  here we have problem q is a matrix now not a vector any more}
Given the caching and demand distributions $\Pm$ and $\Qm$, the asymptotic expected rate is computed as the expected chromatic number of the conflict graph when the number of packets goes to infinity,
%Given $n,m,M$, and the demand distribution $\Qm$, the goal is to find the caching distribution $\Pm$ that minimizes the expected rate 
$\bar R(\Pm,\Qm) \eqdef  \lim_{B \rightarrow \infty} \EE[\chi(\Hsf_{\Csf, \Wsf})/B|\Csf]$,\footnote{$\Hsf_{\Csf, \Wsf}$ denotes the random conflict graph, which is a function of the random caching and demand configurations, $\Csf$ and $\Wsf$, respectively.} where the expectation is taken only over the demand distribution $\Qm$. The expected rate $\bar R(\Pm,\Qm)$ is hence a random variable, which is a function of the random caching configuration $\Csf$.  
We now upper bound $\bar R(\Pm,\Qm)$ using the following Theorem:
\begin{theorem}
\label{thm:up}
For any given $m$, $n$, $M$, and $\Qm$, when $B \rightarrow \infty$, 
the expected rate $\bar R(\Pm,\Qm)$ achieved by a content distribution scheme that uses caching policy in Fig.~\ref{alg1} with caching distribution $\Pm$ %$\{\Pm = [p_{f,u}]: \sum_{f=1}^mp_{f,u} = 1, \forall u; p_{f,u}\leq1/M_{u},\forall f,u\}$, 
and CIC delivery, satisfies
\begin{align}
\label{eq:2}
\bar R(\Pm,\Qm)
\leq \bar R^{\rm ub}(\Pm,\Qm) \eqdef \min\{\psi(\Pm,\Qm),\bar m\},
\end{align}
with high probability.\footnote{The term "with high probability" means that $\lim_{B \rightarrow \infty} \mathbb{P}(\bar R(\Pm,\Qm) \leq \bar R^{\rm ub}(\Pm,\Qm)) = 1$.}
In (\ref{eq:2}), 
\begin{eqnarray}
\label{eq: m bar}
\bar m=\sum_{f=1}^m \left(1 - \prod_{u=1}^n \left(1 - q_{f,u}\right) \right),
\end{eqnarray}
and

\begin{eqnarray}
\label{eq: psi}
\psi(\Pm,\Qm) = \sum_{\ell=1}^n \sum_{\Uc^\ell \subset \Uc}  \sum_{f=1}^m \sum_{u \in \Uc^\ell} 
%\notag\\
%&& 
\rho_{f, u, \Uc^\ell}  \, \lambda(u,f) ,  %\\
\end{eqnarray}
where
\begin{eqnarray}
\lambda(u,f_u) = (1\!-\!p_{f_u,u} \, M_{u})\!\!\!\! \prod_{k \in\Uc^\ell  \setminus \{u\} }\!\!\!\! (p_{f_u,k}M_{k}) \!\!\!\prod_{k \in\Uc 
\setminus \Uc^\ell}\!\!\! (1\!-\!p_{f_u,k}M_{k}), \notag
%&& \qquad  \qquad \prod_{k \in\Uc \setminus \Uc^\ell  }(1-p_{f_u,k}M_{k}),
\end{eqnarray}
 $\Uc^\ell$ denotes a set of users with cardinality $\ell$, and 
\begin{eqnarray}
\rho_{f, u, \Uc^\ell} \eqdef 
\mathbb \PP(f = \arg\! \max_{f_u \in \fv(\Uc^\ell)} \,\,\, \lambda(u,f_u)), \notag%\\
\end{eqnarray}
%\begin{eqnarray}
%\label{eq: psi}
%\psi(\Pm,\Qm) &=& \sum_{\ell=1}^n \sum_{\Uc^\ell \in \Uc}  \sum_{f=1}^m \sum_{u \in \Uc^\ell} \notag\\
%&& \rho_{f, u, \Uc^\ell} (1-p_{f,u} M_u)^{n-\ell+1} (p_{f,u} M_u)^{\ell-1},  \qquad\,
%\end{eqnarray}
%where
%\begin{eqnarray}
%\rho_{f, u, \Uc^\ell} \eqdef \mathbb \PP(f = \underset{f_u \in \fv(\Uc^\ell)}{\text{argmax}} (p_{f,u}M_u)^{\ell-1}(1-p_{f, u}M_u)^{n-\ell+1}) \notag
%\end{eqnarray}
%%&& 
%%$\displaystyle \rho_{f, u, \Uc^\ell} \eqdef %\notag\\
%%%&& 
%%\mathbb \PP(f = \underset{f_u \in \fv(\Uc^\ell)}{\text{argmax}} (p_{f,u}M_u)^{\ell-1}(1-p_{f, u}M_u)^{n-\ell+1})$ %, \notag\\
%%%\end{eqnarray}
denotes the probability that $f$ is the file %whose $p_{f,u}$ 
that maximizes the term $\lambda(u,f)$ among
%{\RED the set of files requested by the subset of $\ell$ users, 
 $\fv(\Uc^\ell)$, the set of files requested by the subset of $\ell$ users $\Uc^\ell$.
\hfill  $\square$
\end{theorem}
%Theorem \ref{thm:up} is proved in Appendix \ref{sec: Proof of Theorem up}.

%\begin{example}

We remark that under homogeneous content popularity and cache size, \ie when $q_{f,u} = q_f, M_u=M, \forall u \in \Uc$, then $p_{f,u} = p_f, \forall u \in \Uc$, and the generalized upper bound of Theorem \ref{thm:up} becomes the order-optimal expected rate for homogenous shared link networks proved in \cite{ji2014average}.

The analytical characterization of the achievable expected rate given by Theorem \ref{thm:up} can then be used to obtain the desired caching distribution for a wide class of heterogeneous network models. %In the following, we refer to the scheme that uses $\Pm^*$ for the caching algorithm in Fig~\ref{alg1} and CIC delivery as {\em Generalized RAndom Popularity-based} (GRAP), with achievable rate $\bar R(\Pm^*,\Qm)$.
In particular, we denote by $\Pm^*$ the caching distribution that minimizes $\bar R^{\rm ub}(\Pm,\Qm)$, and refer to the scheme that uses caching algorithm in Fig.~\ref{alg1} with $\Pm=\Pm^*$ and CIC delivery as RAP-CIC.

%{\RED maybe fig***}

%In \cite{ji2014average}, it shows that RAP, \footnote{In \cite{ji2014average}, we refer the caching scheme with $\Pm^*$ together with the CIC-based coded multicast transmission as {\em RAndom Popularity-based (RAP)}} where $\Pm^*$ is obtained by optimizing (\ref{eq:2 simplified}), is order-optimal and can significantly reduce the average rate compared to other state-of-the-art caching and delivery scheme.
%\hfill $\lozenge$
%\end{example}

%\vspace{-0.1cm}

\section{Polynomial-time Algorithms}
\label{sec: algorithms}

As described in Section \ref{sec: Coded Transmission}, CIC delivery involves computing a minimum vertex coloring of the conflict graph $\mathcal H_{\Cm, \Wm}$. The graph coloring problem is a well known NP-complete problem \cite{GarJoh79}; indeed, given an undirected graph $\mathcal H = (\Vc, \Ec)$, Garey and Johnson showed that obtaining colorings using $s\cdot \chi(\Hc)$ colors, where $s<2$, is NP-hard \cite{GarJoh76a}.

In this section, we describe two polynomial-time delivery schemes. 
The first one is based on a greedy constrained coloring (GCC) algorithm introduced in \cite{ji2014average}.
The authors in \cite{ji2014average} proved that the rate of RAP-GCC converges, as the number of packets $B \rightarrow \infty$ and for homogeneous shared link networks, to the order-optimal expected rate given in Theorem \ref{thm:up} %by $R^{\rm ub}(\Pm,\Qm)=\min\{\psi(\Pm,\Qm),\bar m\}$ in \eqref{eq:2} 
when $q_{f,u} = q_f, M_u=M, \forall u \in \Uc$. 
Following a similar approach, it is immediate to prove that %for heterogeneous shared link networks, given a caching placement as described in Algorithm \ref{alg1}, 
the performance of RAP-GCC in heterogeneous shared link networks converges to the upper bound given in \eqref{eq:2} as $B \rightarrow \infty$.
%In Appendix \ref{sec: Proof of Theorem up}, we prove that when $B \rightarrow \infty$, GCC achieves the upper bound of the average rate for heterogeneous shared link networks given by (\ref{eq:2}). 
It is also easy to verify that RAP-GCC achieves the same performance as the algorithm given in \cite{maddah2013decentralized} for the worst-case demand setting. %In the following we will refer to this algorithm as GCC (Greedy Constrained Coloring).
We recap GCC in Section \ref{sec: An Algorithm achieving upper bound}.

The second one is presented in Section \ref{sec: grasp} and represents a novel coded multicasting scheme based on a Greedy Randomized Algorithm Search Procedure (GRASP) that exhibits lower polynomial-time complexity than GCC. % and is referred to it as GRASP (Greedy Randomized Algorithm Search Procedure). 
In Section \ref{sec: Simulations and Discussions}, we show that for practical regimes of finite file packetization, while GCC loses the multiplicative caching gain, GRASP is still able to approach the fundamental limiting performance and recover a significant part of the multiplicative caching gain.

\vspace{-0.2cm}
\subsection{GCC (Greedy Constrained Coloring)}
% An Algorithm achieving $\bar R^{\rm ub}(\Pm,\Qm)$
\label{sec: An Algorithm achieving upper bound}

%In this section, we recapitulate the GCC which has been introduced in \cite{ji2014average} and is a polynomial-time algorithm  achieving  $\bar R^{\rm ub}(\Pm,\Qm)$ when $F \rightarrow \infty$ and having the same performance as the exponentially (with $n$) complex algorithm in \cite{maddah2013decentralized}. 

The GCC algorithm works by computing two valid colorings of the conflict graph $\Hc_{\Cm,\Wm}=(\mathcal V, \mathcal E)$, referred to as GCC$_1$ and GCC$_2$. GCC compares the rate achieved by the two coloring solutions and constructs the transmission code based on the coloring with minimum rate.\footnote{Recall that the transmission code (index code) is constructed by the modulo sum of the vertices (packets) in $\Hc_{\Cm,\Wm}$ with the same color.}

%\begin{algorithm}[ht]
%\caption{The first coloring algorithm}
%\label{algorithm: coloring 1}
%\begin{algorithmic}[1]
%%\STATE Let ${\mathcal H}_{\Cm, \Wm} = (\Vc, \Ec)$ , where the set of vertices in ${\mathcal H}_{\Cm, \Wm}$ be $\Vc$
%\WHILE{$\Vc \neq \emptyset$}
%\STATE Pick an arbitrary vertex $o$ in $\Vc$. Let $\Ic = \{o\}$.
%\FORALL{$o' \in \Vc/\{o\}$}
%\IF {\{There is no edge between $o'$ and $\Ic$\} $\cap$ \{$\Kc_{o'} = \Kc_{\tilde o}: \forall \tilde o \in \Ic$\}}
%\STATE $\Ic = \Ic \cup o'$.
%\ENDIF
%\ENDFOR
%\STATE Color all the vertices in $\Ic$ by an unused color.
%\STATE $\Vc = \Vc / \Ic$.
%\ENDWHILE
%\end{algorithmic}
%\end{algorithm}
\begin{figure}
\centerline{\fbox{\begin{minipage}[h]{7.5 cm} \normalsize{ %11.8
{\bf algorithm} {\tt GCC$_1$ ($\Vc,\Ec$)} \\
$1$\hspace*{0.2truecm} Let $\hat \Vc = \Vc$; \\
$2$\hspace*{0.2truecm} Let $\Cc = \emptyset$; \\
$3$\hspace*{0.2truecm} $\cv_{\rm 1} = \emptyset$; \\
$4$\hspace*{0.2truecm} {\bf while} $\hat \Vc \neq \emptyset$ \\
$5$\hspace*{0.7truecm} Pick an arbitrary vertex $v$ in $\hat \Vc$. Let $\Ic = \{v\}$; \\
$6$\hspace*{0.7truecm} {\bf for} all $v' \in \hat \Vc/\Ic$$\rightarrow$ \\
$7$\hspace*{1truecm} {\bf if} (There is no edge between $v'$ and $\Ic$\} \\
\hspace*{1.2truecm} $\cap$ \{$\Kc_{v'} \equiv \Kc_{\tilde v}: \forall \tilde v \in \Ic$) {\bf then}  \\
$8$\hspace*{1.2truecm} $\Ic = \Ic \cup v'$; \\
$9$\hspace*{1truecm} {\bf endif}\\
$10$\hspace*{0.5truecm} {\bf endfor} \\
$11$\hspace*{0.5truecm} Color all the vertices in $\Ic$ by $c \notin \Cc$; \\
$12$\hspace*{0.5truecm} Let $\cv_1[\Ic] = c$; \\
$13$\hspace*{0.5truecm} $\hat\Vc = \hat\Vc \setminus \Ic$; \\
$14$\hspace*{0.0truecm} {\bf endwhile}                            \\
$15$\hspace*{0.0truecm} {\bf return}($\cv_{\rm 1}$);          \\
{\bf end} {\tt GCC$_1$} }
\end{minipage}}}
\caption{The greedy constrained coloring algorithm GCC$_1$. %achieving (\ref{eq: psi}) for large enough $B$. 
$\Kc_v$ denotes the set of users that are either caching or requesting packet $v$.}
\label{algorithm: coloring 1}
\vspace{-0.2cm}
\end{figure}

%Let the $\Kc_v$ denote the set of users that are either caching or requesting packet $v$. %(see Example \ref{example: algorithm 1}). 
GCC$_1$ computes a coloring of the conflict graph $\Hc_{\Cm,\Wm}$ as described in Fig.~\ref{algorithm: coloring 1}.  Note that both the outer while-loop starting at line 4 and the inner for-loop starting at line 6 iterate at most $|\Vc|$ times. The operation in line 7 has complexity $O(n)$. Therefore, the complexity of GCC$_1$ is $O(n|\Vc|^2)$ or equivalently $O(n^3B^2)$ since $|\Vc|\leq n B$, which is polynomial in $n, |\Vc|$ (or equivalently in $n,B$).

On the other hand, GCC$_2$ computes a minimum coloring of $\Hc_{\Cm,\Wm}$ subject to the constraint that only the vertices representing the same packet are allowed to have the same color. In this case, the total number of colors is equal to the number of distinct requested packets, and the coloring can be found in $O(|\Vc|^2)$. It is immediate to see that this scheme corresponds to the naive (uncoded) multicasting transmission of all requested packets.  %In another word, we can just transmit this packet such that all the users can receive (decode) simultaneously.

%In Appendix \ref{sec: Proof of Theorem up}, we prove 
It can be shown that RAP-GCC achieves the upper bound of the average rate for heterogeneous shared link networks given by (\ref{eq:2}) (when $B \rightarrow \infty$). However, as will be shown in Section \ref{sec: Simulations and Discussions}, for finite $B$, RAP-GCC loses the promising multiplicative caching gain.

\vspace{-0.1cm}
\subsection{Greedy Randomized Algorithm Search Procedure (GRASP)}
\label{sec: grasp}

%Due to the computational intractability of the Graph Coloring
%Problem,
In this section, we design %and implemented
an efficient metaheuristic to find suboptimal good solutions to the graph coloring problem
%coded multicasting scheme (coloring problem) 
given in Section \ref{sec: Coded
Transmission} in reasonable running times (lower than GCC), and that allow preserving the multiplicative caching gain.

Specifically, we propose a GRASP \cite{FeoRes95a, FesRes02a,FesRes09a,FesRes09b,FesRes11a}, %FeoRes95a (Greedy Randomized Algorithm Search Procedure),
whose general framework is described in the following:
\begin{enumerate}
\item A GRASP performs a certain number of iterations, until a
stopping criterion is met (such as, for example, a predefined maximum number of iterations). %, $\mathtt{MaxIterations}$) %, or  iterations or the a fixed running time);
\item At each GRASP iteration,

\begin{enumerate}
\item a greedy-randomized adaptive solution $\cv$ is built;
\item starting from $\cv$ as initial solution, a local search phase is
performed returning a locally optimal solution $\cv^*$;
\end{enumerate}

\item At the end of all GRASP iterations, the best locally optimal
solution $\cv_{\rm best}$, \ie the solution corresponding to the best objective function value $f(\cv_{\rm best})$ is returned as final
solution and the algorithm stops.
\end{enumerate}

Our GRASP differs from the GRASP proposed by Laguna and Mart\`{\i}
\cite{LagMar01b} in two main aspects: 1) it is able to handle
problem instances characterized by any graph topology,
density/sparsity, and size; 2) the local search strategy checks
for redundant colors focusing on each vertex, one at the time; while
Laguna and Mart\`{\i}'s GRASP
iteratively merges the colors of a pair of independent sets and focuses only on {\em illegal vertices} (\ie
those vertices that, after the merge, result colored with the same
color as one of their adjacent vertices).

Fig.  \ref{f_grasp} depicts the pseudo-code of our GRASP heuristic
for the Graph Coloring Problem. %, where $\Hc_{\Cm, \Wm}=(\Vc,\Ec)$ is an undirected
%graph instance of the problem. $\Vc$ denotes the set of vertices, $\Ec$
%denotes the set of edges, and
%In $\Hc_{\Cm, \Wm}$, for each vertex $i\in \Vc$, $Adj(i)=\{j\in\Vc\ \vert\ [i,j]\in \Ec\}$. 
In the following, we describe the solution construction and the local search procedures performed by our algorithm.

\begin{figure}
\centerline{\fbox{\begin{minipage}[h]{7 cm} \normalsize{ %11.8
{\bf algorithm} {\tt GRASP\_GraphColoring} ($\mathtt{MaxIterations}$, $\Vc$, $\Ec$, $d$, $Adj(\cdot)$, $f(\cdot)$)\\
$1$\hspace*{0.2truecm} $\cv_{\rm best}$ := $\emptyset$; $f(\cv_{\rm best})$ := $+\infty$;  \\
$2$\hspace*{0.2truecm} $\hat \Vc$ := {\tt sort}$(\Vc)$; \\
$3$\hspace*{0.2truecm} {\bf for} $k=1$ {\bf to} $\mathtt{MaxIterations}$$\rightarrow$ \\
$4$\hspace*{0.7truecm} $\Cc$ := $\emptyset$; \\
$5$\hspace*{0.7truecm} $\beta$ := {\tt random}$[0,1]$; \\
$6$\hspace*{0.7truecm} $\cv$ := {\tt BuildGreedyRandAdaptive}($\beta$, $\hat \Vc$, $\Ec$, $d$, $Adj(\cdot)$, $f(\cdot)$, $\Cc$)\\
$7$\hspace*{0.7truecm} $\cv^*$ := {\tt LocalSearch}($\hat \Vc$, $\Ec$, $\cv$, $f(\cv)$, $\Cc$);            \\
$8$\hspace*{0.7truecm} {\bf if} ($f(\cv^*)<f(\cv_{\rm best})$) {\bf then} \\
$9$\hspace*{1truecm} $\cv_{\rm best}$ := $\cv^*$; \\
$10$\hspace*{0.8truecm} $f(\cv_{\rm best})$ := $f(\cv^*)$; \\
$11$\hspace*{0.5truecm} {\bf endif} \\
$12$\hspace*{0.0truecm} {\bf endfor}                            \\
$13$\hspace*{0.0truecm} {\bf return}($\cv_{\rm best}$);          \\
{\bf end} {\tt GRASP\_GraphColoring} }
\end{minipage}}}
\caption{Pseudo-code of a GRASP for the Graph Coloring Problem. $Adj(i)=\{j\in\Vc\ : (i,j)\in \Ec\}, \forall i\in\mathcal V$.}
\label{f_grasp}
\vspace{-0.2cm}
\end{figure}

\textbf{Solution Construction Procedure.}
%Let $\Hc_{\Cm, \Wm}=(\Vc,\Ec)$ be an undirected graph instance of the Graph Coloring
%Problem and
Let $\Qc$ be a set of $|\Vc|$ candidate colors. The construction
phase assigns to each vertex $i\in \Vc$ a color $c \in \Qc$ in such a
way that $c$ is not assigned to any vertex adjacent to $i$
and that the total number of used colors is as smaller as possible.
Fig.~\ref{f_gbuild} %, \ref{f_makeRCL}, and \ref{f_getcolor} 
shows the pseudo-code of the construction procedure. % and the functions therein. % in Fig.~\ref{f_gbuild} are shown in Fig.~\ref{f_makeRCL} and \ref{f_getcolor}. 
To build a feasible solution starting from an empty
solution (line 1), the GRASP construction phase goes through $|\Vc|$ iterations in a for-loop
(lines 2--11), in which colors are assigned to uncolored vertices in a greedy,
randomized, and adaptive manner. In particular, at each iteration,
the choice of the next vertex to be colored is determined by
ordering all currently uncolored vertices in a candidate list $\Wc=\Vc\setminus \cv$ with respect
to a greedy function $g:\,\Wc \mapsto {\mathbb R}$ that measures the
myopic benefit of selecting each vertex and that in our case is
related to the degree of a candidate vertex. The construction is
adaptive because the benefit associated with each candidate vertex
is updated at each iteration of the construction phase to reflect
the changes brought on by the selection of the previous vertex. The
probabilistic component arises from randomly choosing one of
the best candidates in the list $\Wc$, but not necessarily the top
candidate. The list of best candidates is called the {\em Restricted
Candidates List} ($RCL$) and is built according to Fig.~\ref{f_makeRCL}. %We will see later how to build the RCL (see Fig.~\ref{f_makeRCL}).
%selecting the best candidates to be inserted into the list (line 3).
Once the $RCL$ is built (line 3), 
a randomly selected candidate from the $RCL$ (line 4)
is assigned a color based on the colors already assigned to its adjacent vertices (line 5).

\begin{figure}
\centerline{\fbox{\begin{minipage}[h]{7cm} \normalsize{
{\bf function} {\tt BuildGreedyRandAdaptive}($\beta$, $\hat \Vc$, $\Ec$, $d$, $Adj(\cdot)$, $f(\cdot)$, $\Cc$)\\
$1$\hspace*{0.2truecm} $\cv$ := $\emptyset$; \\
$2$\hspace*{0.2truecm} {\bf for} $j=1$ {\bf to} $|\Vc|$$\rightarrow$ \\
$3$\hspace*{0.7truecm} $RCL$ := {\tt MakeRCL} ($\beta$, $\Vc$, $\Ec$, $d$, $\cv$); \\
$4$\hspace*{0.7truecm} $i$ := {\tt SelectIndex} ($RCL$);\\
$5$\hspace*{0.7truecm} $c$ := {\tt GetColor} ($\Vc$, $\Ec$, $i$, $\Cc$, $Adj(\cdot)$, $\cv$);\\
$6$\hspace*{0.7truecm} $\cv[i]$ := $c$;\\
$7$\hspace*{0.7truecm} {\bf if} ($c\notin \Cc$) {\bf then} \\
$8$\hspace*{1truecm} $\Cc$ := $\Cc\cup \{c\}$; \\
$9$\hspace*{1truecm} $f(\cv)$ := $|\Cc|$; \\
$10$\hspace*{0.5truecm} {\bf endif} \\
$11$\hspace*{0.0truecm} {\bf endfor}                            \\
$12$\hspace*{0.0truecm} {\bf return}($\cv$);          \\
{\bf end} {\tt BuildGreedyRandAdaptive} }
\end{minipage}}}
\caption{Pseudo-code of the GRASP construction procedure for the Graph
Coloring Problem.} \label{f_gbuild}
\vspace{-0.3cm}
\end{figure}

%\begin{figure}[htp!]
%\begin{center}
%\subfigure[Scenario 6.i.: $Adj(i)=\{j,k,x,y\}$, $\Cc=\emptyset$. Then,
%$c=$yellow and
%$\Cc=\{\mbox{yellow}\}$.]{\label{f_figure1}\includegraphics[width=30mm]{Figure1}} %40
%\hskip 2em \subfigure[Scenario 6.ii.: $Adj(i)=\{j,k,x,y\}$,
%$\Cc=\{\mbox{yellow,blue}\}$. Then,
%$c=$yellow.]{\label{f_figure2}\includegraphics[width=30mm]{Figure2}}
%\subfigure[Scenario 6.iii.: $Adj(i)=\{j,k,x,y\}$,
%$\Cc=\{\mbox{blue,green,red,gray}\}$. Then, $c=$yellow and
%$\Cc=\Cc\cup\{\mbox{yellow}\}$.]{\label{f_figure3}\includegraphics[width=30mm]{Figure3}}
%\hskip 2em \subfigure[Scenario 6.iv.: $Adj(i)=\{j,k,x,y\}$,
%$\Cc=\{\mbox{blue,green,red,gray}\}$. Then,
%$c=$blue.]{\label{f_figure4}\includegraphics[width=30mm]{Figure4}}
%\end{center}
%\caption{The four possible scenarios that may occur once selected a
%vertex $i$ to be colored during the construction phase.}
%\label{f_build}
%\end{figure}

The construction phase hence performs the following steps (as shown in Figs.~\ref{f_gbuild}, \ref{f_makeRCL}, and \ref{f_getcolor}):

Let $d(i)=|Adj(i)|$, for all $i\in \Vc$, be the degree of vertex $i$.
Let $\cv=\emptyset$ be the solution under construction (initially
empty), \ie the set of vertices already assigned to a color, and let
$\Cc=\emptyset$ (initially empty) be the set of colors that are
associated to at least a vertex in $\cv$. At each iteration, the
following quantities are computed and operations performed:
\begin{enumerate}
\item $g_{\min}$, the minimum greedy value:
\[
g_{\min}=\min_{i\in \Vc\setminus \cv} d(i);
\]

\item $g_{\max}$, the maximum greedy value:
\[
g_{\max}=\max_{i\in \Vc\setminus \cv} d(i);
\]

\item A threshold value $\tau$:
\[
\tau = g_{\min} + \left[\beta \cdot (g_{\max}- g_{\min})\right],\
\mbox{where $\beta\in [0,1]$};
\]

\item The $RCL$, as the subset of candidate uncolored vertices whose degree is at least $\tau$:

\[
RCL=\{i\in \Vc\setminus \cv \ \vert\ d(i)\geq \tau\};
\]

\item A vertex $i$ is randomly selected from the $RCL$ ($i={\tt SelectIndex} ($RCL$)$ in Fig.~\ref{f_gbuild}).

Note that the value of $\beta \in [0,1]$ determines the
percentage of greediness versus randomness in the choice of the
vertices to be inserted in the $RCL$ at each iteration. In fact, for
$\beta=1$, the choice is totally greedy and only vertices with degree
$g_{\max}$ are inserted. On the contrary, for $\beta=0$, the choice
is totally random and all candidate vertices are inserted (\ie
$RCL=\Wc$);

\item Once vertex $i$ is selected, its adjacent vertices are analyzed and
the four possible scenarios that may occur are the following:
\begin{enumerate}
\item[i.] All adjacent vertices are still uncolored and the set $\Cc=\emptyset$: 
in this case, a new color $c$ is assigned to vertex $i$ and
$\Cc=\Cc\cup\{c\}$; %(Fig.~\ref{f_figure1});

\item[ii.] All adjacent vertices are still uncolored and the set $\Cc\not=\emptyset$: 
in this case, vertex $i$ is colored with the first color $c\in \Cc$
available; %(Fig.~\ref{f_figure2});

\item[iii.] At least one adjacent vertex is colored with a color $c\in \Cc$ and all
currently used colors $c\in \Cc$ are already assigned to at least an
adjacent vertex: 
in this case, vertex $i$ is colored with a new color $c'$ and $\Cc=\Cc\cup
\{c'\}$; %(Fig.~\ref{f_figure3});

\item[iv.] At least one adjacent vertex is colored with a color $c\in \Cc$ and there is a color $c'\in
\Cc$ that has not been assigned to any adjacent vertex: 
in this case, vertex $i$ is colored with color $c'$; %(Fig.~
%\ref{f_figure4}).

\end{enumerate}

\item Vertex $i$ is inserted into the solution under construction
 ($\cv[i] = c'$ or $\cv[i] = c$, according to scenarios 6.i.--6.iv.) and the objective function value is updated 
(\ie $f(\cv)=|\Cc|$).
\end{enumerate}

\begin{figure}
\centerline{\fbox{\begin{minipage}[h]{7cm} \normalsize{
{\bf function} {\tt MakeRCL} ($\beta$, $\Vc$, $\Ec$, $d$, $\cv$)\\
$1$\hspace*{0.2truecm} $g_{\min}$ := $\displaystyle{\min_{i\in \Vc\setminus \cv} d(i)}$; \\
$2$\hspace*{0.2truecm} $g_{\max}$ := $\displaystyle{\max_{i\in \Vc\setminus \cv} d(i)}$; \\
$3$\hspace*{0.2truecm} $\tau$ := $g_{\min} + \left[\beta \cdot (g_{\max}- g_{\min})\right]$; \\
$4$\hspace*{0.2truecm} $RCL$ := $\{i\in \Vc\setminus \cv\ \vert\ d(i)\geq \tau\}$; \\
$5$\hspace*{0.2truecm} {\bf return}($RCL$);          \\
{\bf end} {\tt MakeRCL} }
\end{minipage}}}
\caption{Pseudo-code of the function that builds the $RCL$ at each
GRASP construction iteration.} \label{f_makeRCL}
\vspace{-0.2cm}
\end{figure}

\begin{figure}
\centerline{\fbox{\begin{minipage}[h]{7cm} \normalsize{
{\bf function} {\tt GetColor} ($\Vc$, $\Ec$, $i$, $\Cc$, $Adj(\cdot)$, $\cv$)\\
$1$\hspace*{0.2truecm} $\Lc$ := $\emptyset$; \\
$2$\hspace*{0.2truecm} {\bf for each} $j\in Adj(i)$ $\Lc$ := $\Lc\cup \{\cv[i]\}$; \\
$3$\hspace*{0.2truecm} {\bf if} ($\Cc\setminus \Lc\not= \emptyset$) {\bf then} \\
$4$\hspace*{0.7truecm} $c'$ := {\tt SelectColor} ($\Cc\setminus \Lc$); \\
$5$\hspace*{0.2truecm} {\bf else} \\
$6$\hspace*{0.7truecm} $c'$ := {\tt NewColor} ($\Cc$); \\
$7$\hspace*{0.2truecm} {\bf endif} \\
$8$\hspace*{0.2truecm} {\bf return}($c'$);          \\
{\bf end} {\tt GetColor} }
\end{minipage}}}
\caption{Pseudo-code of the function that is invoked at each GRASP
construction iteration to get an available and feasible color.}
\label{f_getcolor}
\vspace{-0.2cm}
\end{figure}

\textbf{Local Search Procedure. }
Solutions generated by the GRASP construction are not guaranteed to
be locally optimal with respect to simple neighborhood definitions.
%{\RED Given a solution $\cv$ and a neighborhood structure $\Nc(\cv)$ %relates $\cv$ to consisting of a subset of solutions close\rq\rq\ to $\cv$, $\cv$ is said to be locally optimal if there is no better solution in $\Nc(\cv)$.} 
It is almost always beneficial to apply a local search to
attempt to improve each constructed solution. A local search
algorithm works in an iterative fashion by successively replacing
the current solution by a better solution in the neighborhood of the
current solution. It terminates when no better solution is found in
the current neighborhood. In our GRASP, the local search algorithm %,
%whose pseudo-code is reported in Fig.~\ref{f_glocal}, 
has the
purpose of checking redundancy of each color $c\in \Cc$, in order to
eventually decrease the current objective function value $|\Cc|$.

%\begin{figure}[htp!]
%\centerline{\fbox{\begin{minipage}[h]{7cm} \normalsize{
%{\bf function} {\tt LocalSearch}($\hat \Vc$, $\Ec$, $\cv$, $f(\cv)$, $\Cc$)\\
%$1$\hspace*{0.2truecm} {\bf for each} $c\in \Cc$$\rightarrow$ \\
%$2$\hspace*{0.7truecm} $\Gc_c$ := $\{i\in V\ \vert\ H[i]=c\}$; \\
%$3$\hspace*{0.7truecm} $\Bc$:=$\emptyset$; \\
%$4$\hspace*{0.7truecm} $\hat \cv$ := $\cv$; \\
%$5$\hspace*{0.7truecm} {\bf for each} $i\in \Gc_c$$\rightarrow$ \\
%$6$\hspace*{1truecm} $\Ac$ := $\emptyset$; \\
%$7$\hspace*{1truecm} {\bf for each} $j\in Adj(i)$ $\Ac$ := $\Ac\cup \{\cv[j]\}$; \\
%$8$\hspace*{1truecm} {\bf if} ($\Cc\setminus \Ac \not= \emptyset$) {\bf then} \\
%$9$\hspace*{1.5truecm} $c'$ := {\tt SelectColor} ($\Cc\setminus \Ac$); \\
%$10$\hspace*{1.3truecm} $\hat \cv[i]$ := $c'$; \\
%$11$\hspace*{1.3truecm} $\Bc$ := $\Bc\cup \{i\}$; \\
%$12$\hspace*{0.8truecm}  {\bf endif} \\
%$13$\hspace*{0.5truecm} {\bf endfor} \\
%$14$\hspace*{0.5truecm} {\bf if} ($|\Bc|$ = $|\Gc_c|$) {\bf then} \\
%$15$\hspace*{0.8truecm}  $\cv$ := $\hat \cv$; \\
%$16$\hspace*{0.8truecm}  $\Cc$ := $\Cc\setminus \{c\}$; \\
%$17$\hspace*{0.8truecm}  $f(\cv)$ := $|\Cc|$; \\
%$18$\hspace*{0.5truecm} {\bf endif} \\
%$19$\hspace*{0.0truecm} {\bf endfor}                            \\
%$20$\hspace*{0.0truecm} {\bf return}($\cv$);          \\
%{\bf end} {\tt LocalSearch} }
%\end{minipage}}}
%\caption{Pseudo-code of the GRASP local search for the Graph
%Coloring Problem.} \label{f_glocal}
%%\vspace{-0.5cm}
%\end{figure}

In more detail, %in loop for in lines 1--19 in Fig.~\ref{f_glocal}, 
the local search computes, 
iteratively for each color $c\in \Cc$, the set $\Gc_c$ of all
vertices colored with color $c$ and performs the following
steps:

\begin{enumerate}
\item For each vertex $i\in \Gc_c$, span $Adj(i)$: %, i.e., the vertices adjacent to vertex $i$: 
if there is a color $c'\in \Cc$, $c'\not= c$, not assigned to any adjacent vertex
$j\in Adj(i)$, then color vertex $i$ with color $c'$;

\item Color $c$ is removed from the set $\Cc$ if and only if in Step 1 it
has been possible to replace $c$ %associated with each vertex $i\in \Gc_c$ 
with some color $c'\not= c$.

\end{enumerate}

\textbf{Computational complexity of the proposed GRASP. }
%Given the input undirected graph $\Hc_{\Cm,\Wm}=(\Vc,\Ec)$, 
The complexity analysis of the proposed GRASP algorithm boils down to the following steps: 
%to solve the Graph Coloring Problem of is as follows:
\begin{enumerate}
\item Sorting the set $\Vc$ according to a non-ascending order of the degree of the vertices:  %hasa computational time equal to %:
%\[
$O(|\Vc|\, \log|\Vc|)$.
%\]
\item Solution construction procedure: %hase runs $|\Vc|$ iterations andat each iteration it performs the following steps:
$O(|\Ec|)$.
%\begin{itemize}
%\item construction of the $RCL$:
%since $g_{\min}$, $g_{\max}$, and $\tau$ can be computed in $O(1)$
%(the vertices are sorted according to their degree), to identify the
%candidate vertices to be inserted in the $RCL$ (according to the rule
%described in Section \ref{s-build}) requires $O(|\Vc|)$;
%
%\item color assignment:
%to this end, it must be scanned the adjacent vertices of vertex $i$
%randomly extracted from the $RCL$ in order to choose the color that
%will be assigned to it in an appropriate way. 
%This task requires $O(|Adj(i)|)$. Therefore, summing all over vertices
%$i\in \Vc$ to be colored, we obtain $\displaystyle{\sum_{i=1}^{|\Vc|}
%|Adj(i)| = 2 \cdot |\Ec|}$.
%It is clear that the computational complexity of the construction
%procedure is equal to %:
%%\[
%$O(|\Ec|)$.
%%\]
%
%\end{itemize}
%
\item Local Search Procedure:
%For each used color $c\in \Cc$ and for each $i\in \Gc_c$, the local search
%procedure analyzes the adjacent vertices of vertex $i$ in order to
%attempt to remove color $c$ from the set $\Cc$.
%Since $|\Cc|+|\Gc_c|=O(|\Vc|)$, this procedure has a computational
%complexity equal to %:
%\[
$O(|\Ec|)$.
%\]

\end{enumerate}

It can be seen that step 1) is performed only once, at the beginning of the algorithm.
Since step 2) and step 3) are performed a fixed number of
iterations ($\mathtt{MaxIterations}$), it results that the overall
computational complexity of the proposed GRASP %algorithm to solve the Graph Coloring problem 
has computational complexity
\vspace{-0.1cm}
%{\small 
\begin{eqnarray}
%&& O(|\Vc|\,\log|\Vc| + \mathtt{MaxIterations} \cdot \max\{|\Vc|,|\Ec|\}) \notag\\
&& O(|\Vc|\,\log|\Vc| + \mathtt{MaxIterations} \cdot |\Ec|).
\end{eqnarray}
%}

{\em Remark:} Observe that the complexity of GRASP is $O(|\Vc|^2)$, which is a factor of $n$ lower than the complexity ($O(n|\Vc|^2)$) of GCC. %the algorithm achieving $\bar R^{\rm ub}(\Pm,\Qm)$ for large enough $B$. %shown in Fig. \ref{algorithm: coloring 1}.

%%%%%%%%%%%%%%%%%%%%%%%%%%%%%%%%%%%%%%%%%%%%%%%%%%%%%%%%%%
\begin{figure*}[ht]
%\vspace{-3cm}
\centering
%\hspace{-1.3cm}
\subfigure[]{
%\vspace{-5cm}
%\hspace{-1cm}
\centering \includegraphics[width=4.65cm, height=4cm]{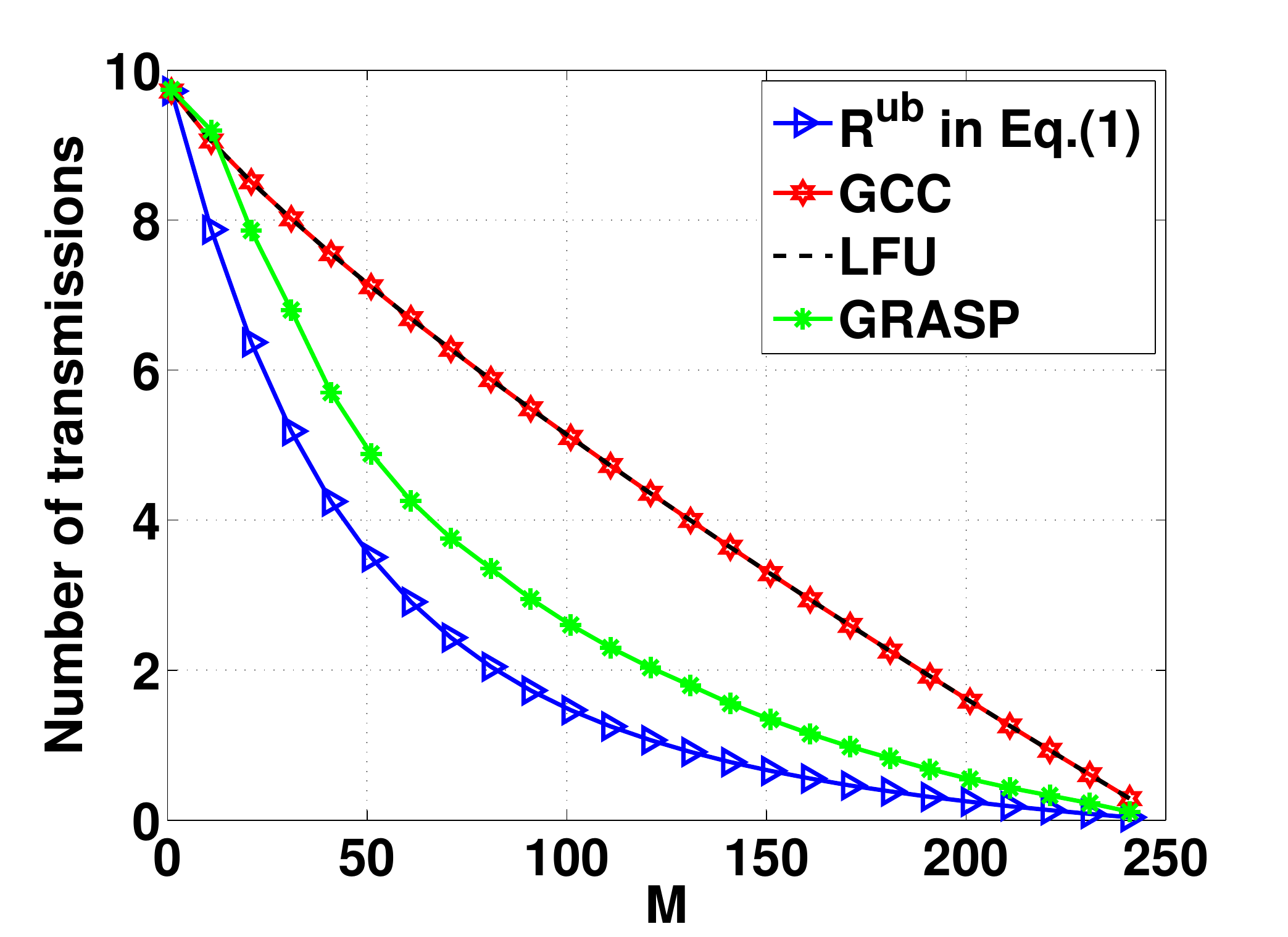}
%\vspace{-5cm}
%\hspace{-3cm}
\label{fig1020}
}
\hspace{-0.7cm}
\subfigure[]{
%\vspace{-5cm}
%\hspace{-1cm}
\centering \includegraphics[width=4.65cm, height=4cm]{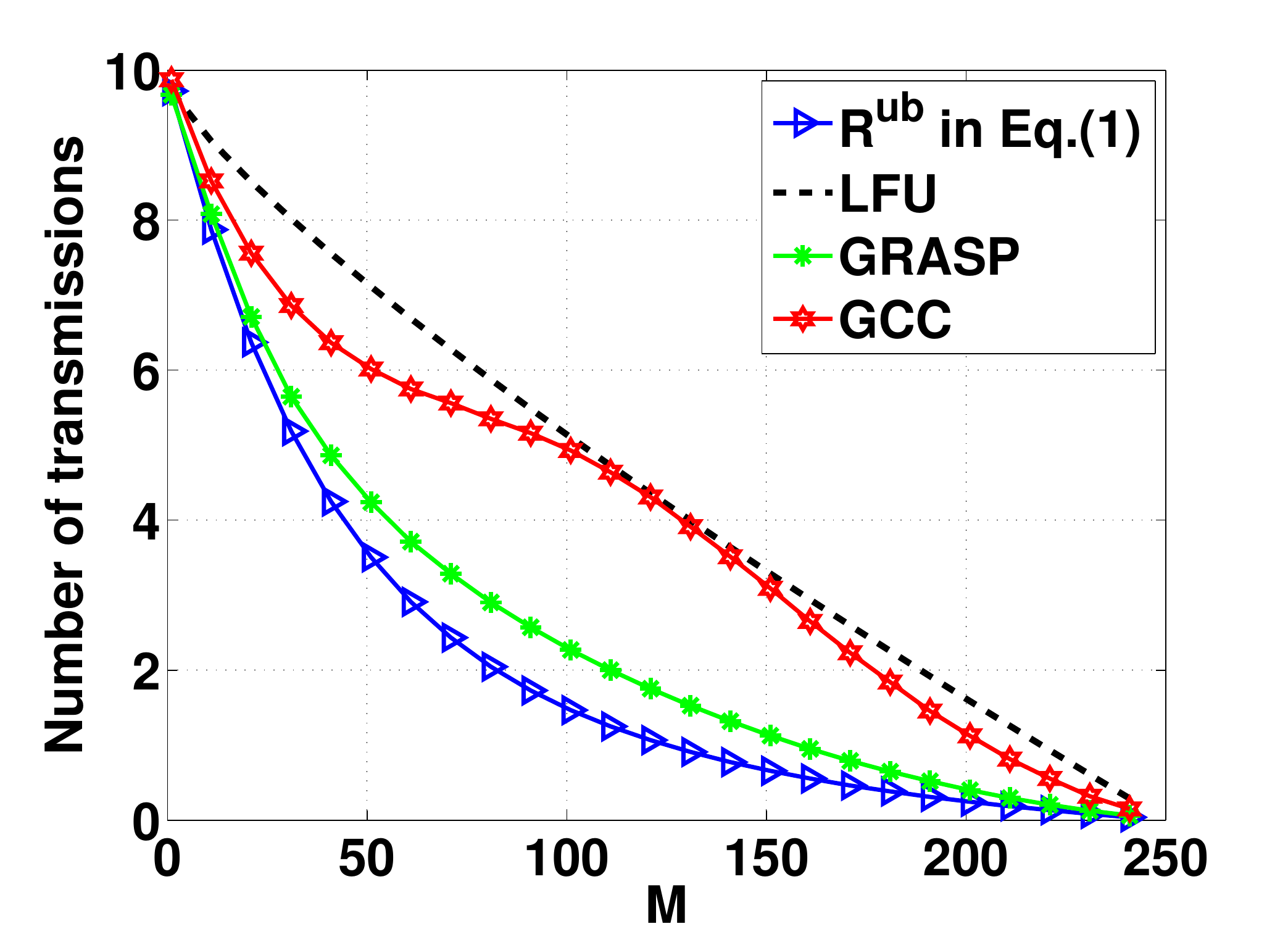}
%\vspace{-5cm}
%\hspace{-3cm}
\label{fig10100}
}
\hspace{-0.7cm}
\subfigure[]{
%\vspace{-5cm}
%\hspace{-0.8cm}
\centering \includegraphics[width=4.65cm, height=4cm]{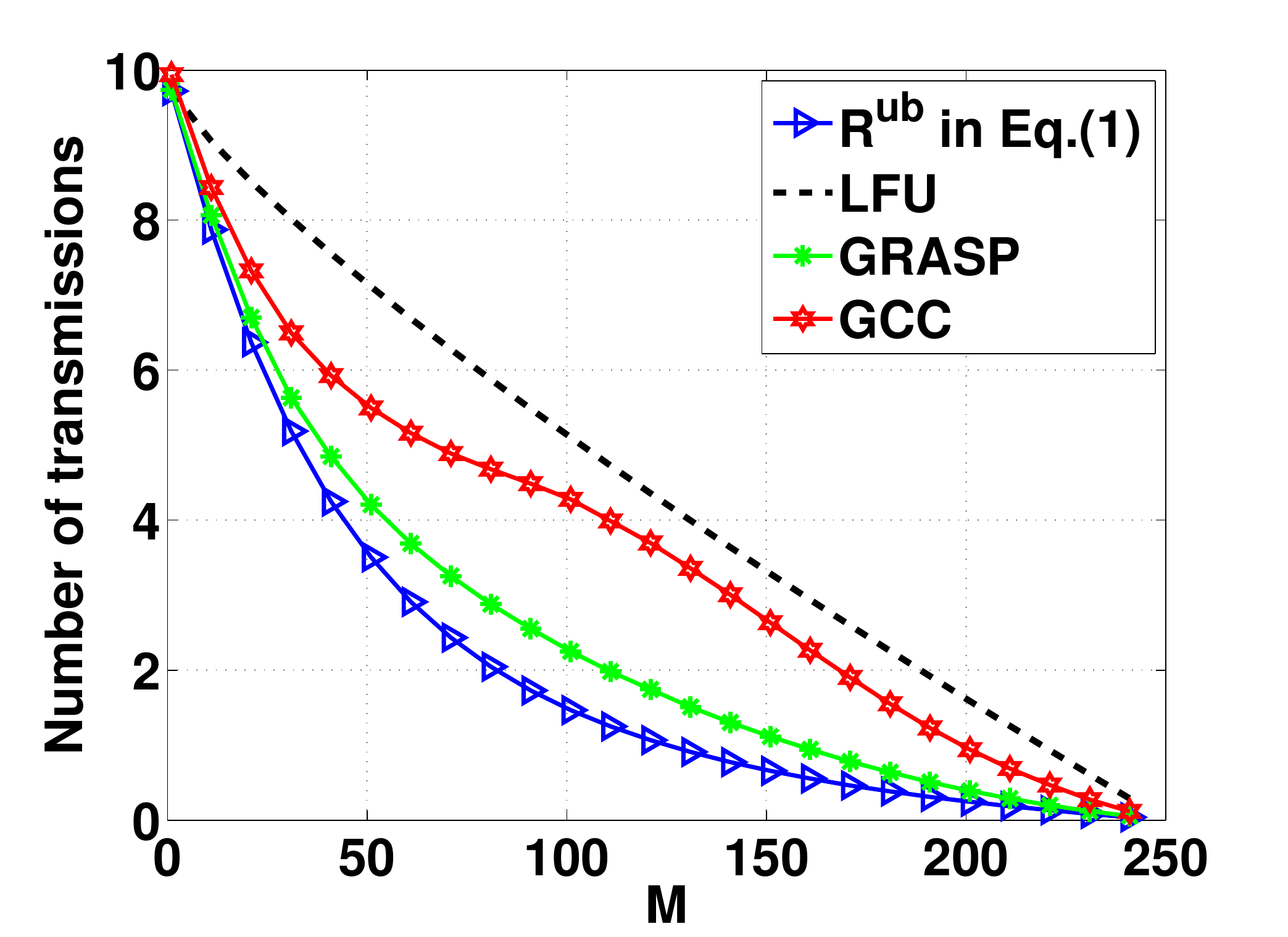}
%\vspace{-5cm}
%\hspace{-3cm}
\label{fig10200}
}
\hspace{-0.7cm}
\subfigure[]{
%\vspace{-5cm}
%\hspace{-1cm}
\centering \includegraphics[width=4.65cm, height=4cm]{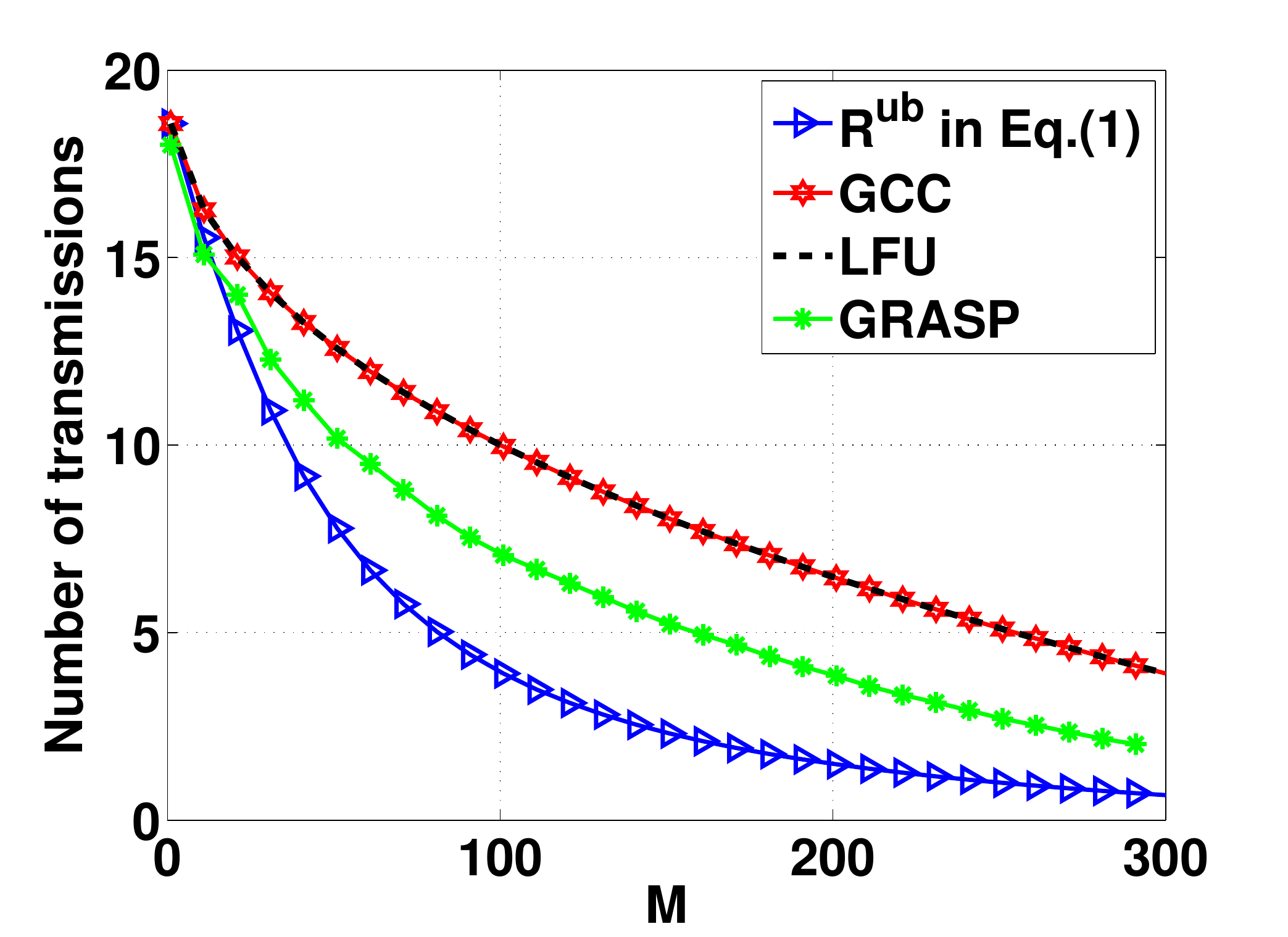}
%\vspace{-5cm}
%\hspace{-3cm}
\label{fig2050}
}
%\vspace{-0.1cm}
\caption{Average number of transmissions over the shared multicast link. a)~$n=10$, $m= 250$, $B=20$, and $\alpha=0.2$; b)~$n=10$, $m= 250$, $B=100$, and  $\alpha=0.2$; c)~$n=10$, $m= 250$, $B=200$, and $\alpha=0.2$; d)~$n=20$, $m= 500$, $B=50$, and  $\alpha=0.6$.}
\vspace{-0.5cm}
\label{fig: result 1}
\end{figure*}

\vspace{-0.2cm}
\section{Simulations and Discussions}
\label{sec: Simulations and Discussions}

%As already mentioned in the previous subsections in  \cite{ji2014average},  \cite{maddah2012fundamental}, and \cite{maddah2013decentralized}  schemes exploiting the combined benefit of network coding and in-network cooperative caching capabilities have been proposed and shown to be able to achieve significant gains in caching network with a shared multicast link under the assumption of infinite packetizations. 
In this section, we numerically analyze the performance of the two polynomial-time achievable schemes illustrated in Section \ref{sec: algorithms} for finite file packetization. Specifically, assuming the random popularity-based (RAP) caching policy in Fig. \ref{alg1} with caching distribution $\Pm=\Pm^*$ (recall that $\Pm^*$ is the caching distribution that minimizes $\bar R^{\rm ub}(\Pm,\Qm)$ in (\ref{eq:2}) among all $\Pm$), we compare the average performance of GCC and GRASP when files are partitioned into a finite number of packets $B$.  
For comparison, we also plot 1) the performance of LFU (Least Frequently Used),\footnote{LFU discards the least frequently requested file upon the arrival of a new file to a full cache of size $M_u$ files. In the long run, this is equivalent to caching the $M_u$ most popular files.}  shown to be optimal in uncoded networks, and  2) the performance of GCC for infinite file packetization ($B\rightarrow \infty$), as given in Theorem \ref{thm:up}.

For simplicity and to illustrate the effectiveness of GRASP, we consider a homogeneous network scenario, in which users request files according to a Zipf demand distribution with parameter $\alpha \in \{0.2,0.6\}$ and all caches have size $M$ files. 
%For all considered schemes, the RAP caching distribution $\Pm^*$  is obtained by minimizing $\bar R^{\rm ub}(\Pm,\Qm)$ in (\ref{eq:2}) among all $\Pm$. % described by a $m$-dimensional vector taking value in $ \{\frac{1}{\tilde{m}}, 0\}$.\footnote{This constraint on the caching distribution introduced in \cite{ji2014average},  originates a scheme referred to as Random Least-Frequently-Used (Random LFU), which approximates RAP and generalizes the well known LFU caching scheme. In Random LFU, each user just caches packets from the (carefully designed) $\tilde{m}$ most popular files in a distributed and random manner.},
%{\BLUE {\RED We refer GCC as the two algorithm achieving the upper bound before, so GCC definitely includes LFU. We may want to rewrite this BLUE sentences.} 

Further, we assume that when using GCC or GRASP, %based on the system parameters (number of users, $n$, number of files, $m$, number of packets, $B$ and Zip parameter $\alpha$),  
the source node pre-evaluates the performance of LFU and chooses the minimum of the two accordingly. %to adopt between LFU and  GCC, the scheme minimizing the result rate over the shared link.  Similarly assumption is done for GRASP. 
Hence, denoting by $R_{LFU}$, $R_{GCC}$ and $R_{GRASP}$ the average rate achieved by LFU, GCC and GRASP, respectively, Fig. \ref{fig: result 1} plots the performance of GCC and GRASP as  $\min\{R_{LFU}, R_{GCC}\}$, and $\min\{R_{LFU}, R_{GRASP}\}$, respectively.\footnote{Note that LFU may be slightly better than GRASP only for small $B$ and very small $M$ (negligible caching benefit), as shown in Fig. \ref{fig: result 1}(a).}

Figs. \ref{fig: result 1}(a), (b), and (c) %\ref{fig1020}, \ref{fig10100} and \ref{fig10200} %\ref{fig: result 1} %a, b and c 
plot the average rate for a network with $n=10$ users, $m=250$ files and Zipf parameter $\alpha=0.2$. Observe how the significant caching gains (with respect to LFU) quantified by the order-optimal upper bound are completely lost when using GCC with finite packetization $B=20$, and only slightly recovered as the packetization increases to $B=100$ and $B=200$. On the other hand, observe how GRASP remarkably preserves most of the promising multiplicative caching gains for the same values of file packetization. 
For example, in Fig.~\ref{fig10100}, if $M$ doubles from $M=50$ to $M=100$, then the rate achieved by GRASP essentially halves from $4.2$ to $2.2$.
%we can see that GRASP achieves an average rate of $4.2$ and if $M$ is doubled, the achievable average rate by GRASP becomes $2.2$, which is almost half of rate for $M=50$. 
For the same regime, it is straightforward to verify that neither GCC nor LFU exhibits this property. %\footnote{Note that additive and multiplicative gains may show indistinguishable when $M$ is comparable to the library size $m$. Hence, one needs to pick a reasonably small $M$ ($\frac{m}{n} < M \ll m$) to observe the multiplicative caching gain of GRASP.} 
Fig. \ref{fig: result 1}(d) illustrates a scenario with higher popularity skewness, e.g., $\alpha=0.6$. Observe how, also in this setting, a finite number of packets ($B=50$) completely limits the gains of GCC. On the other hand, GRASP is still able to preserve significant gains. For example, when $M$ doubles from $M=70$ to $M=140$, the achievable rate by GRASP goes from $8.8$ to $5.5$, approaching a half rate reduction even with only $50$ packets per file.  
Finally note from Fig. \ref{fig: result 1}(b), that in oder to guarantee a rate $R=4$, GCC requires a cache size of $M=120$, while GRASP can reduce the cache size requirement to $M=50$, a $2.4\times$ cache size reduction. 

\bibliographystyle{IEEEbib}
\bibliography{references,references_d2d}

\end{document}